\def\year{2020}\relax
\documentclass[letterpaper]{article} % DO NOT CHANGE THIS
\usepackage{aaai20}  % DO NOT CHANGE THIS
\usepackage{times}  % DO NOT CHANGE THIS
\usepackage{helvet} % DO NOT CHANGE THIS
\usepackage{courier}  % DO NOT CHANGE THIS
\usepackage[hyphens]{url}  % DO NOT CHANGE THIS
\usepackage{graphicx} % DO NOT CHANGE THIS
\usepackage{graphicx}  % DO NOT CHANGE THIS
\usepackage{multirow}
\usepackage{subcaption}
\usepackage{amsmath,amssymb,amsfonts}
\usepackage{color}

\title{Reinforcement-Learning based Portfolio Management with Augmented Asset Movement Prediction States}
\author{Yunan Ye,\textsuperscript{\rm 1}
Hengzhi Pei,\textsuperscript{\rm 2}
Boxin Wang,\textsuperscript{\rm 3}
Pin-Yu Chen,\textsuperscript{\rm 4}
Yada Zhu,\textsuperscript{\rm 4}
Jun Xiao,\textsuperscript{\rm 1}
Bo Li\textsuperscript{\rm 3}\\
\textsuperscript{\rm 1}Zhejiang University, 
\textsuperscript{\rm 2}Fudan University, 
\textsuperscript{\rm 3}University of Illinois at Urbana-Champaign, 
\textsuperscript{\rm 4}IBM Research\\
chryleo@zju.edu.cn, hzpei16@fudan.edu.cn, boxinw2@illinois.edu,\\ pin-yu.chen@ibm.com, yzhu@us.ibm.com, junx@zju.edu.cn, lbo@illinois.edu
}
\begin{document}
\maketitle
\begin{abstract}

Portfolio management (PM) is a fundamental financial planning task that aims to achieve investment goals such as maximal profits or minimal risks. Its decision process involves continuous derivation of valuable information from various data sources and sequential decision optimization, which is a prospective research direction for reinforcement learning (RL). 
In this paper, we propose SARL, a novel State-Augmented RL framework for PM. 
Our framework aims to address two unique challenges in financial PM: (1) \textit{data heterogeneity} -- the collected information for each asset is usually diverse, noisy and imbalanced (e.g., news articles); and (2) \textit{environment uncertainty} -- the financial market is versatile and non-stationary. To incorporate heterogeneous data and enhance robustness against environment uncertainty, our SARL augments the asset information with their price movement prediction as additional states, where the prediction can be solely based on financial data (e.g., asset prices) or derived from alternative sources such as news. Experiments on two real-world datasets, (i) Bitcoin market and (ii) HighTech stock market with 7-year Reuters news articles, validate the effectiveness of SARL over existing PM approaches, both in terms of accumulated profits and risk-adjusted profits. 
Moreover, extensive simulations are conducted to demonstrate the importance of our proposed state augmentation, providing new insights and boosting performance significantly over standard RL-based PM method and other baselines.

\end{abstract}

\section{Introduction}

An investment portfolio is a basket of assets that can hold stocks, bonds, cash and more. In today's market, an investor's success heavily relies on maintaining a well-balanced portfolio. Despite the recent trend of big data and machine intelligence have triggered resolution to the investment industry, portfolio management is still largely based on linear models and the Markowitz framework  \cite{Papenbrock2016}, known as Modern Portfolio Theory (MPT). 
MPT relies on accurate prediction of market prices and restricted assumptions such as past probability distribution of assets’ returns fully representing the future. Accurate market price forecast is extremely challenging, if not impossible, due to its highly noisy, stochastic and chaotic nature \cite{Tsay2010}. Essentially, PM involves sequential decision making of continuously reallocating a number of funds into assets based on the latest information to achieve the investment goal. It is natural to leverage reinforcement learning (RL) to model the decision making process for asset reallocation~\cite{almahdi2017adaptive,jiang2017deep,liang2018adversarial}. Although some attempts have made to apply RL for financial PM,  there are still several challenges remain unsolved: 1) \emph{data heterogeneity}: the collected information for each product may be sparse, noisy, imbalanced, and diverse (e.g., financial time series vs unstructured data, news articles), and therefore it is difficult to incorporate different information within the single model. 2) \emph{Environment uncertainty}: the non-stationary nature of financial markets induces uncertainty and often causes a distribution shift between training and testing data.

To address the aforementioned challenges, in this paper, we propose a novel State Augmented RL (SARL)  framework for PM. The proposed SARL aims to leverage additional diverse information from alternative sources other than classical structured financial data, such as asset prices, to make market trend prediction. Such prediction will then be incorporated into our RL framework for state augmentation. SARL is a general framework in the sense that it can incorporate different sources of information into the augmented states. It is also not restricted to any particular market. Throughout this paper, the information sources mainly include the asset price, the exemplary financial data, and news articles of different products (companies), the exemplary alternative data.

To evaluate the performance of SARL, we test it on two real-world datasets: the Bitcoin cryptocurrency dataset ~\cite{jiang2017deep} and the HighTech stock dataset~\cite{ding2014using}.
As an example, SARL is able to achieve $140.9\%$ improvement over the state-of-the-art method~\cite{jiang2017deep} in terms of accumulative return on the Bitcoin dataset.
SARL also outperforms all PM baselines including CRP, OLMAR, WMAMR in terms of portfolio value and the Sharpe ratio~\cite{sharpe1964theory}.
In addition, we show that due to market uncertainty, the standard RL based PM approaches are difficult to generalize while SARL is much more generalizable and easier to be adaptive to new environments. For instance, SARL achieves $11.73\%$ and $40.45\%$ improvement in Sharpe Ratio (SR) on Bitcoin and HighTech datasets respectively.
Our contributions are listed as follows:
\begin{itemize}
    \item We propose a novel State Augmented RL framework (SARL) for PM, where additional information from different sources can be leveraged to make market prediction and such prediction will be embedded as augmented states in the RL framework to improve PM performance.
    \item We evaluate the proposed SARL on two real-world datasets and show that SARL outperforms the state-of-the-art method by $140.9\%$ on the Bitcoin dataset and $15.7\%$ on the HighTech dataset.
    \item We perform in-depth analysis for the standard RL based PM approach and SARL and show that SARL is more generalizable than the former regarding data distribution shift. We also conduct simulations for justification. 
    \item We conduct extensive experiments and find that: 1) The exploitation of diverse information can reduce the impact of environment uncertainty. 2) We show that high-density (more frequent) external information can help boost the overall PM performance even when the information is noisy. 3) Low-density but high-accuracy external information can also improve the PM performance. 

\end{itemize}

\begin{figure*}[htb]
    \centering
    \includegraphics[width=1.6\columnwidth]{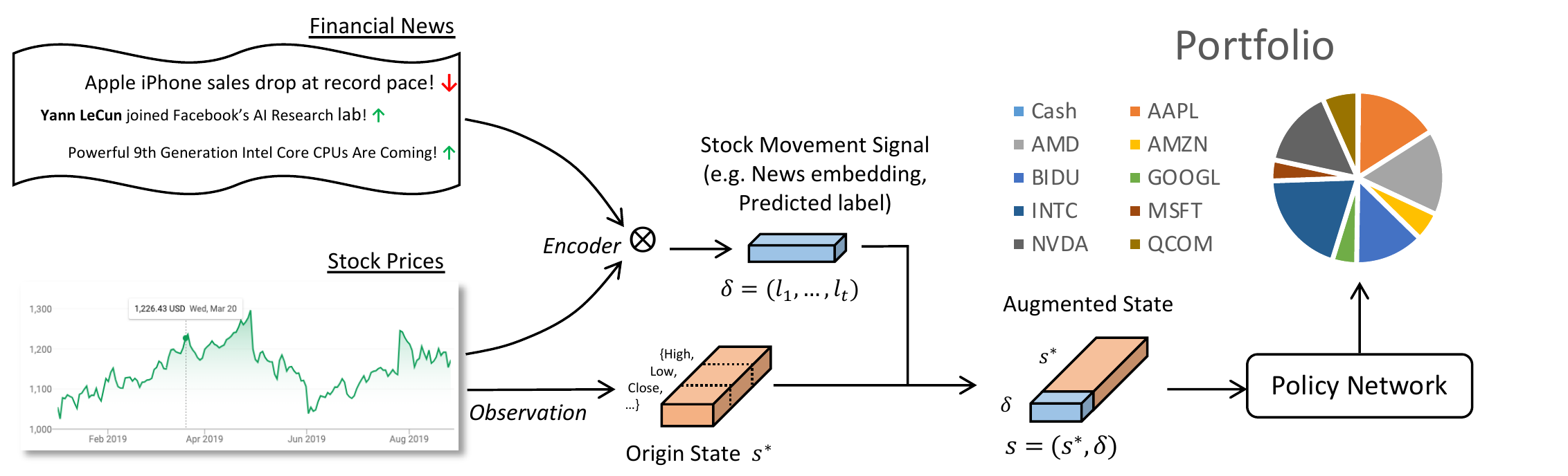}
    \caption{The framework of our proposed State Augmented Reinforcement Learning (SARL) method. Asset prices (e.g., stock prices) represent the internal features constituting the original state $s^*$, and  Financial News represent the external information to be encoded and augmented to the final state $s$ for SARL. The asset movement signal $\delta$ is used for state augmentation and the policy network will generate the portfolio management strategy from the augmented state $s$.}
    \label{fig:SARL}
\end{figure*}

\section{Related Work}

With the availability of large scale market data, it's natural to employ deep learning (DL) model which can exploit the potential laws of market in PM. Prior arts~\cite{heaton2017deep,schumaker2012evaluating,nguyen2015sentiment} in training a neural network (NN) model for market behavior prediction have shown their effectiveness in asset price prediction and asset allocation. However, DL models which have no interaction with the market has a natural disadvantage in decision making problem like PM. Reinforcement learning algorithms have been proved effective in decision making problems in recent years and deep reinforcement learning (DRL)~\cite{chen2019counterfactual}, the integration of DL and RL, is widely used in the financial field.
For instance, \cite{almahdi2017adaptive} proposed a recurrent reinforcement learning (RRL) method, with a coherent risk-adjusted performance objective function named the Calmar ratio, to obtain both buy and sell signals and asset allocation weights. \cite{jiang2017deep} use the model-free Deep Deterministic Policy Gradient (DDPG) \cite{lillicrap2015continuous} to dynamically optimize cryptocurrency portfolios. Similarly, \cite{liang2018adversarial} optimize asset portfolios by using the DDPG as well as the Proximal Policy Optimization (PPO)~\cite{schulman2017proximal}. \cite{buehler2019deep} presents a DRL framework to hedge a portfolio of derivatives under transaction costs, where the framework does not depend on specific market dynamics. However, they mainly tackle the PM problem by directly utilizing the direct observation of historical prices for RL training, which may largely overlook data noise and overestimate the model's learning capability.

\section{Background and Problem Formulation}
\label{section:PF}
Portfolio management (PM) is a fundamental financial planning task that aims to maximize forecasted profits (or minimize calculated risks) via asset allocation. A market is made up of many assets and their related information, e.g. prices and other factors that affect the market. We assume the market is sufficiently liquid such that any transactions can be executed immediately with minimal market impact.
For PM, we consider the scenario that there's a machine learning algorithm that can gather all viable information from the market and then gradually improves its trading strategy by trial-and-error. Here the market consisting of all the assets for PM and other available information is called the \textbf{environment}.
Based on the liquidity hypothesis, the algorithm which observes the environment and then makes decisions to interact with the market and rebalance the portfolio can be defined as an \textbf{agent}. 

Throughout this paper, we consider the setup that the environment will provide asset prices as an internal data source and will also provide financial news articles (when available) as an external data source. The agent has access to all historical prices and news articles up to the current time step for making low-level predictions such as price changes or high-level predictions such as asset movements (up/down). Intuitively, an agent that gives accurate asset price change predictions is ideal, but it is hard to be trained in practice due to market uncertainties and possible distribution shifts between training (past market) and testing (future market) environments. On the other hand, predicting high-level changes such as asset movements may be an easier task, which in turn gives a more reliable predictive information when augmented with the asset prices for reallocating portfolios. 

Let $v_{i,t}$, $i=\{1,\dots,n\}$ denote the closing price of the $i^{th}$ asset at time $t$, where $n$ is the number of assets to be considered for PM.  The price vector $\boldsymbol{v_t}$ consists of the closing prices of all $n$ assets. Similarly, $\boldsymbol{v^H_t}$ and $\boldsymbol{v^L_t}$ denote the highest prices and the lowest prices at time step $t$, respectively. For instance, $t$ is an index of asset trading days. It is worth noting that in PM problems, the assets are not always fully invested. 

In addition to the portfolio of $n$ assets, we introduce an additional dimension (the first dimension indexed by $0$) in $\boldsymbol{v_t}$, $v_{0,t}$, to denote the ``cash price'' at time instance $t$. 
As we normalize all temporal variations in $\boldsymbol{v_t}$ with respect to cash value, $v_{0,t}$ remains constant for all $t$.

By modeling the PM problem as a Markovian decision process which indicates that the next state only depends on current state and action. We can formulate the PM problem as a triplet $(S,A,r)$, where $S$ is a set of states, $A$ is a set of actions, and $r:S\times A\rightarrow\mathbb{R}$ is the reward function. 

To be akin to asset price changes over time, we denote $\boldsymbol{y_t}=\frac{\boldsymbol{v_{t+1}}}{\boldsymbol{v_t}}$ as the relative price vector. More precisely, 
\begin{equation}
\boldsymbol{y_t}=\frac{\boldsymbol{v_{t+1}}}{\boldsymbol{v_t}}=(1,\frac{v_{1,t+1}}{v_{1,t}},\dots,\frac{v_{n,t+1}}{v_{n,t}})^T
\end{equation}
To formulate the process of asset reallocation in PM,
we introduce the reallocation weight fraction $\boldsymbol{w_t}=(w_{0,t},w_{1,t},\dots,w_{n,t})^T$ in our framework. $w_{i,t}, t\ne0$ is the weight fraction of the $i^{th}$ asset and $w_{0,t}$ is the weight fraction of cash at the end of time step $t$. The asserts are reallocated based on the weight assigned to each asset. Thus, we have $\sum_{i=0}^n w_{i,t}=1$. 

In our implementation, the RL agent sells or buys assets based on the difference of $\boldsymbol{w_t}$ and $\boldsymbol{w_{t-1}}$ between time steps $t-1$ and $t$ to reallocate the portfolio.

\section{SARL: A Framework of Deep RL with Augmented Asset Movement Prediction States}

As we formulate the PM problem as a decision making process with heterogeneous data, we propose a hierarchical approach which binds supervised learning and RL into a unified
framework such that it can be trained with standard RL methods, as illustrated in Figure~\ref{fig:SARL}. We employ an end-to-end network to extract asset movement information from either internal source (e.g., Price Up/Down Predicted label from historical prices) or external source (e.g., News embedding). We then integrate it with the prices of assets for state augmentation. Finally, we adopt a deterministic policy gradient algorithm based on the augmented state for learning the policy of PM. We note that our SARL framework can easily incorporate different data sources through the use of an encoder for state augmentation.

\subsection{Augmented Asset Movement Prediction State}

It is common in practical PM methods to integrate human-engineered features with asset prices for better prediction and asset reallocation. In our SARL framework, we have made it generic to be capable of incorporating heterogeneous data into the standard RL training pipeline. As illustrated in Figure \ref{fig:SARL}, we propose an \textit{encoder} $\delta$ that takes different types of data sources and transforms their contents into informative representations to be augmented to the asset prices for training an RL agent. For example, the \textit{encoder} $\delta$ can be a classifier that takes past asset prices over a certain period as inputs and produces asset movement predictions, or it can be a feature extraction function derived from a text classifier, which is trained on the word embeddings of news for asset movement prediction. Overall, the augmented state is defined as 

\begin{equation}
s=(s^*,\delta) 
\end{equation} 
where $s^*$ is the observable state (i.e., current asset prices) related to low-level observations and $\delta$ is the encoder summarizing high-level observations (i.e., asset movement prediction from past asset prices or news).

In our SARL framework, we offer the flexibility to either adopt the internal or the external information to augment the state. For \textbf{internal information}, we use past prices to predict the asset movement and then integrate the prediction result into the state. The intuition is that in PM problems, the asset price is the most critical information. Augmenting asset movement prediction from past asset prices can offer some additional high-level and robust information to improve decision making, which may not be apparent or sufficiently expressed when merely using raw asset prices for RL.

Specifically, we train a recurrent neural network with long short-term memory (LSTM)~\cite{hochreiter1997long} to predict the asset movement. The binary output (price up/down) will guide the model to choose a better strategy. For \textbf{external feature}, we collect financial news articles related to the assets selected for PM since they provide new but possibly correlated information for asset movement prediction in addition to asset prices. We use different kinds of Natural Language Processing (NLP) methods as encoders to embed the news and then feed the embedding into a hierarchical attention network (HAN)~\cite{yang2016hierarchical} to train a binary classifier to predict the price movement. The features in the last layer before the softmax layer are extracted to represent the embedding of the news. Finally, we integrate the embedding into the state for augmentation. One advantage of our state-augmented RL is
its generality in incorporating heterogeneous data sources via encoders and its compatibility with standard RL training algorithms via state augmentation. As will be evident in the Experiments section, state augmentation with either internal or external information indeed significantly improves the performance of PM.

\subsection{Deterministic Policy Gradient}
Deterministic Policy Gradient (DPG) learns a deterministic target policy using deep neural networks. A policy is a mapping from the state space to the action space, $\pi_{\theta}: S \rightarrow A$. Policy gradient represents the policy by a parametric probability distribution $\pi_{\theta}(a|s)=P(a|s;\theta)$ that selects action $a$ from action space in state $s$ according to parameter vector $\theta$. For a deterministic policy $a=\mu_{\theta}(s)$, the selection is deterministically produced by the policy from a state. Since the return $r_{t}^{\gamma}$ is defined as the total discounted reward from time-step $t$ onwards reward, $r_{t}^{\gamma} = \sum_{k=t}^{\infty}\gamma^{k-t}r(s_k,a_k)$ where $r$ is the reward function and $\gamma$ is the discount factor where $0<\gamma<1$. We define the performance objective as $J(\mu_{\theta}) = \mathbb{E}[r_{1}^{\gamma}|\mu]$, 
which is the expectation over the discounted state distribution $\rho^{\mu}(s)$ defined as

\begin{equation}
\begin{aligned}
J(\mu_{\theta})&=\int_{S}\rho^{\mu}(s)r(s,\mu_{\theta}(s))\mathrm{d}s\\
&=\mathbb{E}_{s\sim\rho^{\mu}}[r(s,\mu_{\theta}(s))]
\end{aligned}
\end{equation}

Considering the time interval from $1$ to $T$, the corresponding performance objective function is 
\begin{equation}
\begin{aligned}
J_{T}(\mu_{\theta})=\sum_{t=1}^T\gamma^{t}r(s_t,\mu_{\theta}(s_t))
\end{aligned}
\label{eq:eq4}
\end{equation}

The objective in \eqref{eq:eq4} is the typical Markovian decision process objective function. However, we note that this type of function doesn't match the portfolio management task due to the property that the assert accumulated by time $t$ would be reallocated in time $t+1$~\cite{jiang2017deep,liang2018adversarial,kanwar2019deep}. We follow~\cite{liang2018adversarial} to modify the objective function, which makes it more suitable for the portfolio management task. To be specific, we replace the summation by the accumulated product of the portfolio value $P_T=\prod_{t=1}^{T}P_{0}r_t$. Thus, the performance objective becomes
\begin{equation}
J_{T}(\mu_{\theta})=J_0\prod_{t=1}^{T}r(s_t,\mu_{\theta}(s_t)),
\end{equation}
where $J_0$ is a constant. 

\subsection{Action and Reward for Portfolio Optimization}
\textbf{Action($a$)}. As we illustrated before, we use fraction vector of the total assets at time step $t$, $\boldsymbol{w_t}=\{w_{0,t},w_{1,t},\dots,w_{n,t}\}^T$ to represent the allocation of the assets. What the agent should do is to reallocate the assets into assets, that is, adjust $\boldsymbol{w_{t+1}}$. The desired reallocating weights at time step $t$, $\boldsymbol{a_t}=\{a_{0,t}, a_{1,t},\dots,a_{n,t}\}^T$, with the constraint $\sum_{i=0}^n a_{i,t}=1$, is the action vector in our model. By taking the action at time step $t$, the asset allocation vector would be influenced by the price movement $\boldsymbol{y_t}$. At the end of the time period, the allocation vector $\boldsymbol{w_{t+1}}$ becomes
\begin{equation}
\boldsymbol{w_{t+1}}=\frac{\boldsymbol{y_t}{\odot}\boldsymbol{a_t}}{\boldsymbol{y_t}\cdot\boldsymbol{a_t}}
\end{equation}
where $\odot$ is the element-wise multiplication. 

\textbf{Reward($r$)}. The reward function of each time step can be defined in a standard way based on profit the agent made. The fluctuation of the value of the assets for each asset is $a_{t,i}{\cdot}y_{t,i}$. Thus the total profit at time step $t$ is $\boldsymbol{a_{t}}\cdot\boldsymbol{y_t}$. Taking transaction cost $\beta=\sum_{i=1}^{n}|a_{i,t}-w_{i,t}|$ into consideration, the immediate reward at time step $t$ can be represented as:
\begin{equation}
r_t=r(s_t,a_t)=\boldsymbol{a_{t}}\cdot\boldsymbol{y_t}-\beta\sum_{i=1}^{n}|a_{i,t}-w_{i,t}|
\end{equation}
The introduction of transaction cost makes the formulation close to the real market operation but brings more difficulty in mathematical formalism. Obviously, this reward function is not differentiable in this form. Applying the approximation of $\beta$ in~\cite{jiang2017deep} which approximate $\beta_t$ with portfolio vectors of two recent periods and the price movement vector, we get $\beta_t=\beta_t(\boldsymbol{w_{t-1}},\boldsymbol{w_t},\boldsymbol{y_t})$. A fixed constant commission rate for non-cash assets trading is applied. We set $c_b=c_s=0.25\%$ where $c_b$ and $c_s$ is the constant commission rate of \textit{buy} and \textit{sell}. Follow the setting of the modified DPG, the target of the agent is to maximize the accumulated product value, which is equivalent to maximize the sum of the logarithmic value. Finally we get the modified reward function at time step $t$:
\begin{equation}
r_t=r(s_t,a_t)=\ln{(\beta_t\boldsymbol{a_{t}}\cdot\boldsymbol{y_t})}
\end{equation}
and the accumulative return $R$:
\begin{equation}
\begin{aligned}
R(s_1,a_1,\dots,s_T,a_T)=J_T=\frac{1}{T}\sum_{t=1}^T\ln{(\beta_t\boldsymbol{a_t{\cdot}y_t})}
\end{aligned}
\end{equation}

Considering the policy $\mu_\theta$, our goal is to maximize the objective function parameterized by $\theta$, we can formally write it as below:

\begin{equation}
\begin{aligned}
\mu_{\theta{*}}&=\arg\max_{\mu_\theta} J_T(\mu_\theta)
\\&=\arg\max_{\mu_\theta} \frac{1}{T}\sum_{t=1}^T\ln{(\beta_t\boldsymbol{a_t{\cdot}y_t})}
\end{aligned}
\end{equation}

\begin{equation}
\nabla_{\theta}\mu_{\theta}(\tau)= \frac{1}{T}\sum_{t=1}^T \nabla_{\theta} \ln{\mu_{\theta}(a_t,s_t)}
\end{equation}
\begin{equation}
\theta\leftarrow\theta+\lambda\nabla_{\theta}\mu_{\theta}(\tau)
\end{equation}
where $\lambda$ is the learning rate. Due to the existence of the denominator $T$, the equation is properly normalized for data with different length $T$, which also makes mini-batch training over a sampled time period feasible. 

\section{Experimental Results}

In this section, we compare the performance of SARL with other methods on two datasets: Bitcoin and HighTech. We will summarize these two datasets, specify their data challenges, define the evaluation metrics, introduce the baseline PM methods for comparison and perform extensive experiments and simulations to validate the importance of state augmentation in SARL for PM.

\subsection{Datasets}
We use the following two datasets from different markets. 

\noindent\textbf{Bitcoin}~\cite{jiang2017deep} contains the prices of 10 different cryptocurrencies from \textit{2015-06-30} to \textit{2017-06-30}. For every cryptocurrency, we have $35089$ data points representing prices recorded in a half-hour interval. To evaluate the performance, we split the dataset into training ($32313$ data point) and testing ($2776$ data point) parts chronologically.

\begin{figure}[!h]
    \centering
    \includegraphics[width=0.7\columnwidth]{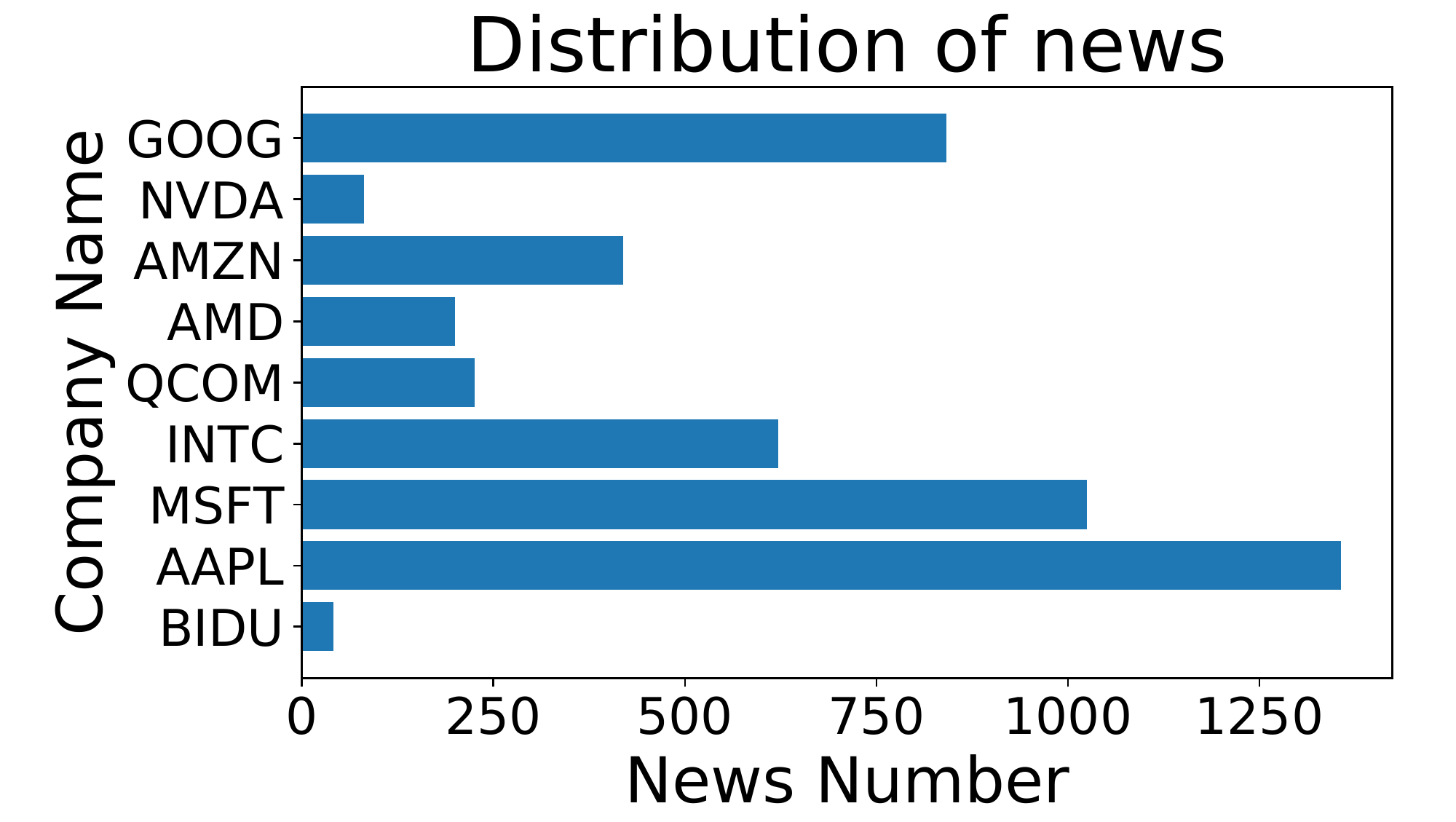}
    \caption{The distribution of news in HighTech dataset.}
    \label{fig:distribution}
\end{figure}

\noindent\textbf{HighTech}~\cite{ding2014using} consists of both daily closing asset price and financial news from \textit{2006-10-20} to \textit{2013-11-20}. Even we select the companies with as much news as possible, the distribution of the news articles of chosen companies in Figure~\ref{fig:distribution} shows that we can't have everyday news for a particular company. In order to have sufficient external information to justify the use of state augmentation in a real market, we choose 9 companies which have the most news from our dataset. There are in total 4810 articles related to the $9$ stocks filtered by company name matching. These news articles cover $72.48\%~(1293/1784)$ of the trading days in this dataset. We split the dataset into $1529$ days for training and $255$ days for testing chronologically.

\subsection{Evaluation Metrics}
We report two widely-used metrics for measuring the performance of a particular PM strategy. The most intuitive metric is to analyze the accumulative change of assets' values over the test time, which is called Portfolio Value (PV). 
To take the risk factor into consideration, we also report the Sharpe Ratio (SR), risk-adjusted return.

\begin{itemize}
    \item \textbf{Portfolio Value} Given any PM strategy, the assets' Portfolio Value (PV)~\cite{ormos2013performance} at the final time horizon $T$ can be calculated by:
    \begin{equation}
        {
        p_{T} =p_{0} \exp \left(\sum_{t=1}^{T} r_{t}\right)=p_{0} \prod_{t=1}^{T} \beta_{t} \boldsymbol{a}_{t} \cdot \boldsymbol{y}_{t}
        }   
    \end{equation} 
    PV is used to quantify the amount of profit the strategy earns in time length $T$ as it is an accumulative value. Based on the liquidity assumption, without loss of generality, we normalize the portfolio with starting value $p_0=1$. 
    \item \textbf{Sharpe Ratio} The Sharpe Ratio~\cite{sharpe1964theory} is the average return earned in excess of the risk-free rate per unit of volatility or total risk. It is used to compare the portfolio's return to its risk and is defined as 
    \begin{equation}
        \textbf{Sharpe Ratio} = \frac{R_{p}-R_{f}}{\sigma_{p}}
    \end{equation}
    where $R_p$ is the return of the portfolio, $R_f$ is the risk-free rate and $\sigma_p$ is the standard deviation of the portfolio's excess return. Here we use $R_p=\sum_{t=1}^{T} \beta_{t} \boldsymbol{a}_{t} \cdot \boldsymbol{y}_{t}$ and set $R_f=2\%$ as a typical bank interest value.
\end{itemize}

\subsection{Baselines}
\begin{itemize}
    \item \textbf{CRP}. Constant rebalanced portfolio (CRP)~\cite{cover2011universal} is an investment strategy which keeps the same distribution of wealth among a set of assets from day to day. That is, the proportion of total wealth in a given asset is the same at the beginning of each day.
    \item \textbf{OLMAR}. On-Line Moving Average Reversion (OLMAR)~\cite{li2012line} is a method that exploits moving average reversion to overcomes the limitation of single-period mean reversion assumption. 
    \item \textbf{WMAMR}. Weighted Moving Average Mean Reversion(WMAMR)~\cite{gao2013weighted} is a method which fully exploits past period price relatives using equal-weighted moving averages and then learns portfolios by online learning techniques.
    \item \textbf{EW}. Equal Weight is a naive baseline which lets the agent uniformly invest all the stocks. The tendency of EW can reflect the macroscopic movement of the market.
    \item \textbf{DPM}. Deep Portfolio Management~\cite{jiang2017deep} is a method based on Ensemble of Identical Independent Evaluators (EIIE) topology. DPM uses asset prices as state and trains an agent with a Deep Neural Network (DNN) approximated policy function to evaluate each asset's potential growth in the immediate future and it is the state-of-art RL algorithm for PM.
\end{itemize}

\subsection{Data Challenge}
Before we compare different PM strategies on the Bitcoin and HighTech datasets, we address their data challenges including unbalanced data distribution, noisy data and environment uncertainty, which will be used to explain the performance variations of different PM methods.
%%%sparsity

\textit{\textbf{Unbalanced data distribution}} -- Although we already selected the 9 companies having the most news for performance evaluation,  Figure~\ref{fig:distribution} shows the distribution of financial news is highly unbalanced across companies. For instance, Apple has $33$ times more news than Baidu. 
In addition, the unbalanced news distribution also occurs across time -- each company is not guaranteed to have a related news article every day. 
More generally, not every state $s_t$ will have its corresponding external feature. 

    \begin{figure}[t]
    \centering
    \begin{subfigure}[b]{0.9\columnwidth}
        \centering
        \includegraphics[width=\columnwidth]{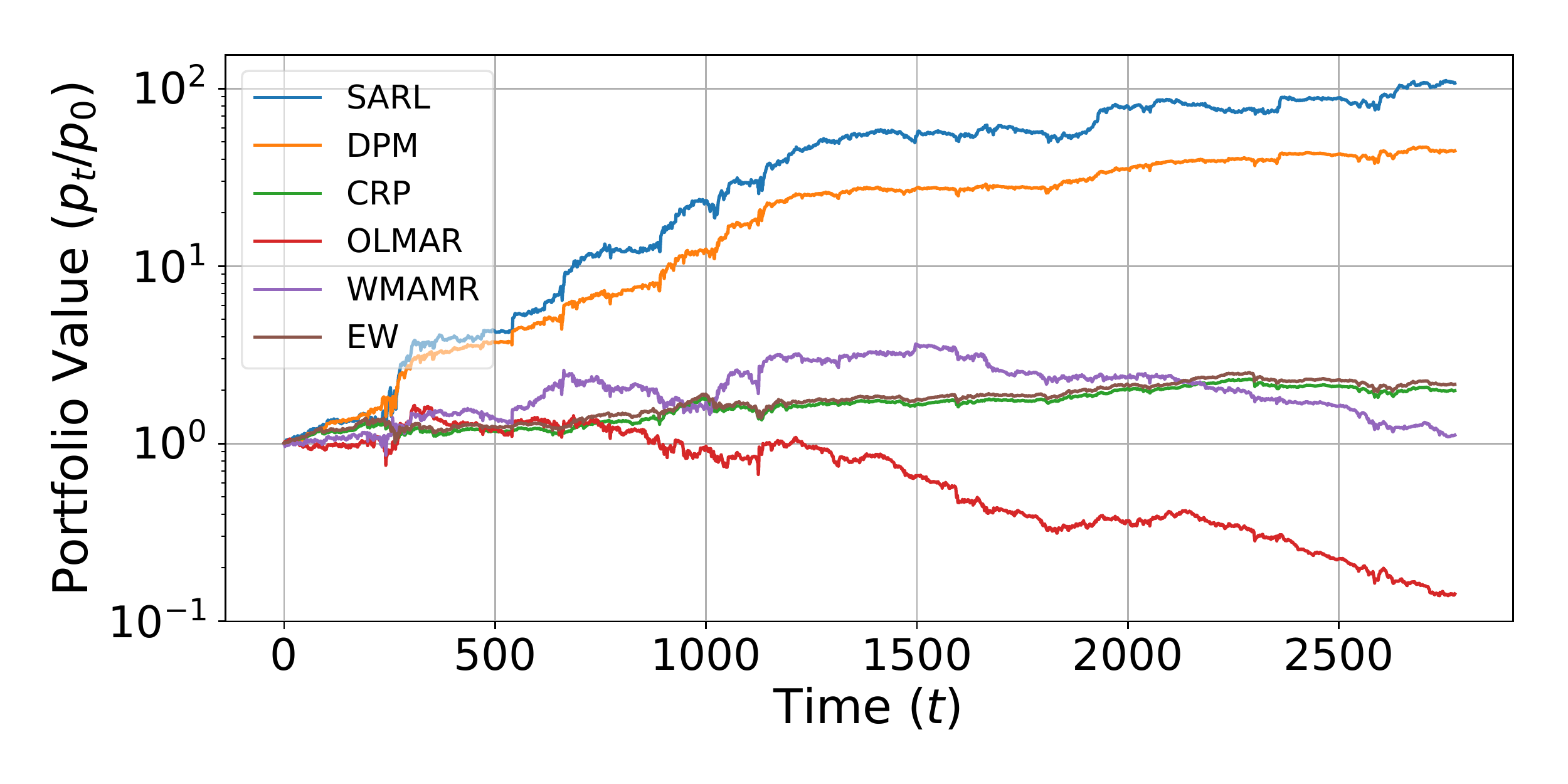}
        \caption{Portfolio Value of different methods on Bitcoin dataset.}
        \label{fig:bitcoin_best}
    \end{subfigure}
    \hfill
    \begin{subfigure}[b]{0.9\columnwidth}
        \centering
        \includegraphics[width=\columnwidth]{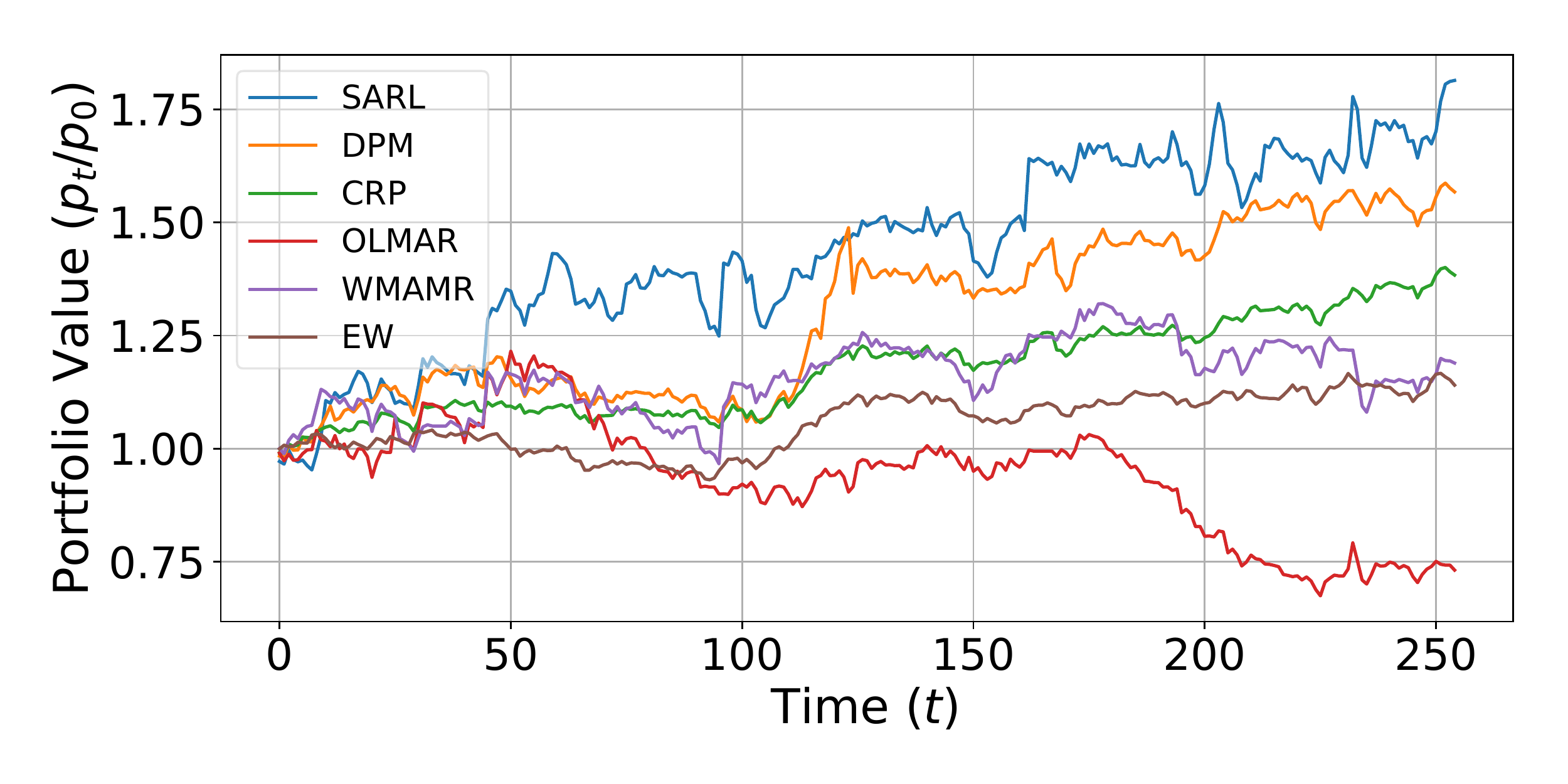}
        \caption{Portfolio Value of different methods on HighTech dataset.}
        \label{fig:reuters_best}
    \end{subfigure}
    \caption{The portfolio value of different PM methods. SARL is our proposed state augmentation RL method, DPM is the state-of-art standard RL method in PM. CRP, OLMAR, WMAMR are baseline financial PM methods.}
    \end{figure}

%%%correlation
\textit{\textbf{Noisy external data}} -- Since news can cover multiple topics and we keep a news article as long as a company's name is mentioned regardless of the context, there could be a lot of redundant and useless information which is irrelevant to portfolio management.
To justify the issue of inherent noise in news, we train a text classifier hierarchical attention network (HAN)~\cite{yang2016hierarchical} on three different word embedding techniques, including Glove~\cite{pennington2014glove}, Word2Vec~\cite{mikolov2013distributed} and Fasttext~\cite{joulin2016bag}. We also adopt Auto Phrase~\cite{shang2018automated} for phrase mining and train a random initialized embedding network. The average training and testing accuracies for stock movement prediction are reported in Table~\ref{tab:NLP}, where the best training/testing accuracy is no greater than 66\%/61\%, suggesting noisy information in the collected news.

\begin{table}[!h]
\centering
\caption{The training and testing accuracy of the text classifier for different word embedding methods.}
\label{tab:NLP}
\resizebox{1\columnwidth}{!}{
\begin{tabular}{c|cccc}
\hline
\textbf{Method} & \textbf{Auto Phrase} & \textbf{Glove} & \textbf{Word2Vec} & \textbf{Fasttext} \\ \hline
\textbf{Train}  & 0.6592               & 0.6410          & 0.6508             & 0.6207            \\ \hline
\textbf{Test}   & 0.6022               & \textbf{0.6099} & 0.6061             & 0.5908            \\ \hline
\end{tabular}
}

\end{table}

%%%sharpe ratio
\textit{\textbf{Environment uncertainty}} -- Different from game playing tasks which have certain rules, PM is deeply influenced by the market dynamics. The PM strategy of a standard RL agent trained on past market dynamics may not be generalizable to the future market if there are substantial changes in market dynamics (i.e., the problem of distribution shift). We will use the Sharpe ratio of different testing time periods to illustrate the influence of environment uncertainty. 

\subsection{Results and Discussion}
Here we report the results on Bitcoin and HighTech datasets and discuss the performance of SARL and other baselines.
    \begin{table}[t]
    \centering
    \caption{Sharpe Ratio of different time period in Bitcoin dataset.\protect\footnotemark (w:week, m:month)}
    \label{tab:Bitcoin}
    \resizebox{0.7\columnwidth}{!}{
    \begin{tabular}{c|cccc}
    \hline
    \multirow{2}{*}{\textbf{Method}} & \multicolumn{4}{c}{\textbf{Sharpe Ratio}}                    \\ \cline{2-5} 
                                     & \textbf{1w}    & \textbf{2w}    & \textbf{1m}    & \textbf{2m}    \\ \hline
    \textbf{CRP}                     & 4.70           & 3.00           & 4.82           & 3.95           \\ \hline
    \textbf{OLMAR}                   & 6.78           & 2.99           & 0.75          & -2.62           \\ \hline
    \textbf{WMAMR}                   & 5.33           & 6.93           & 4.88           & 1.30           \\ \hline
    \textbf{EW}                     & 5.32           & 3.68           & 4.77           & 3.95           \\ \hline
    \textbf{DPM}                          & 14.25          & 13.37          & 12.64          & 9.80           \\ \hline
    \textbf{SARL}                  & \textbf{14.78} & \textbf{14.61} & \textbf{13.89} & \textbf{10.60} \\ \hline
    \end{tabular}
    }
    \end{table}
    
    \begin{table}[t]
    \centering
    \caption{Sharpe Ratio of different time period in HighTech dataset. (w:week, m:month)}
    \label{tab:Reuters}
    \resizebox{0.65\columnwidth}{!}{
    \begin{tabular}{c|ccccc}
    \hline
    \multirow{2}{*}{\textbf{Method}} & \multicolumn{4}{c}{\textbf{Sharpe Ratio}}    \\ \cline{2-5} 
                                     & \textbf{2w}       & \textbf{1m}       & \textbf{3m}       & \textbf{6m}       \\ \hline
    \textbf{CRP}            & 6.92          & 3.44          & 2.49          & 2.06          \\ \hline
    \textbf{OLMAR}          & 1.51          & 0.41          & 1.61          & -0.05          \\ \hline
    \textbf{WMAMR}          & 6.61          & 0.42          & 1.74          & 1.30          \\ \hline
    \textbf{EW}             & 4.69          & 0.74          & 0.13          & 1.54          \\ \hline
    \textbf{DPM}            & 6.44          & \textbf{3.88}          & 2.31          & 2.22          \\ \hline
    \textbf{SARL}           & \textbf{7.73} & 3.83 & \textbf{2.91} & \textbf{2.37} \\ \hline
    \end{tabular}
    }
    \end{table} 
\footnotetext{In Bitcoin dataset, 48 data points are included in a day. }
 \textbf{Portfolio Value (PV).} In Bitcoin dataset, we use the previous prices of the past 30 days to train a classifier for price up/down prediction. We employ a neural network based on LSTM as an encoder and the classifier has $65.08\%$ training and $60.10\%$ testing accuracy. In HighTech dataset, we use the financial news related to stocks for classifier training. We choose Glove as the embedding method and employ a HAN as an encoder to obtain a 100-dimensional embedding vector of stock movement prediction for each news. The training/testing accuracy is $64.10\%$/$60.99\%$. The division of training and testing of the classifier of each dataset follows the aforementioned setting.
    In our SARL training, we use the prices of past 30 days as standard state $s^*$. In Bitcoin, we use their related prediction labels for state augmentation. In HighTech, the average news embeddings of past 12 days are used for state augmentation. In case there is no external source at time $t$, we set $\delta$ to be zero.
    The PV at the final horizon in testing set as shown in Figure \ref{fig:bitcoin_best} and \ref{fig:reuters_best} demonstrates the effectiveness of SARL. SARL improves PV by $140.9\%$ and $15.7\%$ when compared to the state-of-art RL algorithm for PM (DPM) in Bitcoin and HighTech respectively. Since the market of cryptocurrency has more volatility, it's easier for a good agent to gain more profits. 
    
    %% figure of pv
    \begin{figure}[t]
    \centering
    \begin{subfigure}[b]{0.48\columnwidth}
        \centering
        \includegraphics[width=\columnwidth]{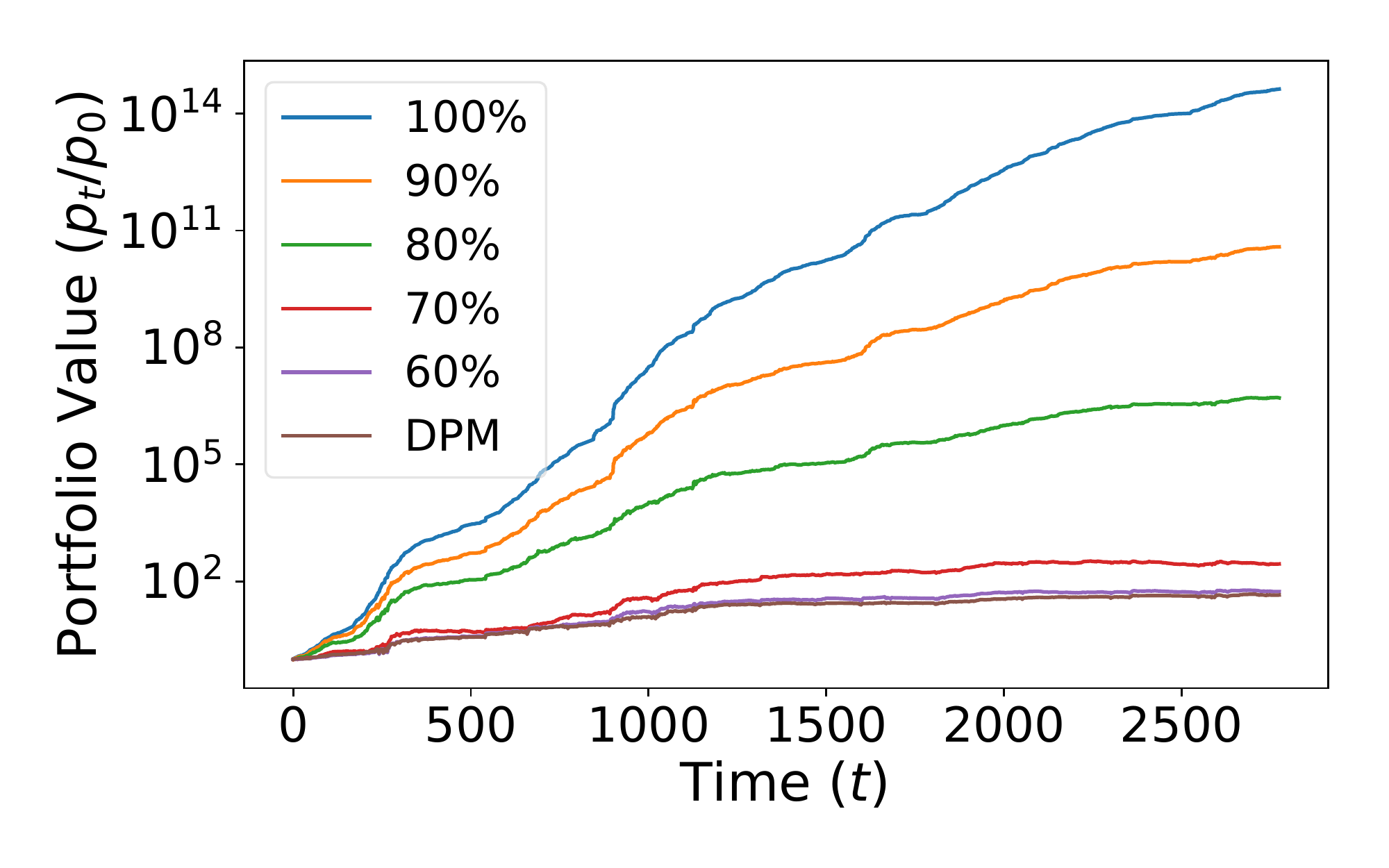}
        \caption{Bitcoin simulation}
        \label{fig:bitcoin_simulation}
    \end{subfigure}
    \hfill
    \begin{subfigure}[b]{0.48\columnwidth}
        \centering
        \includegraphics[width=\columnwidth]{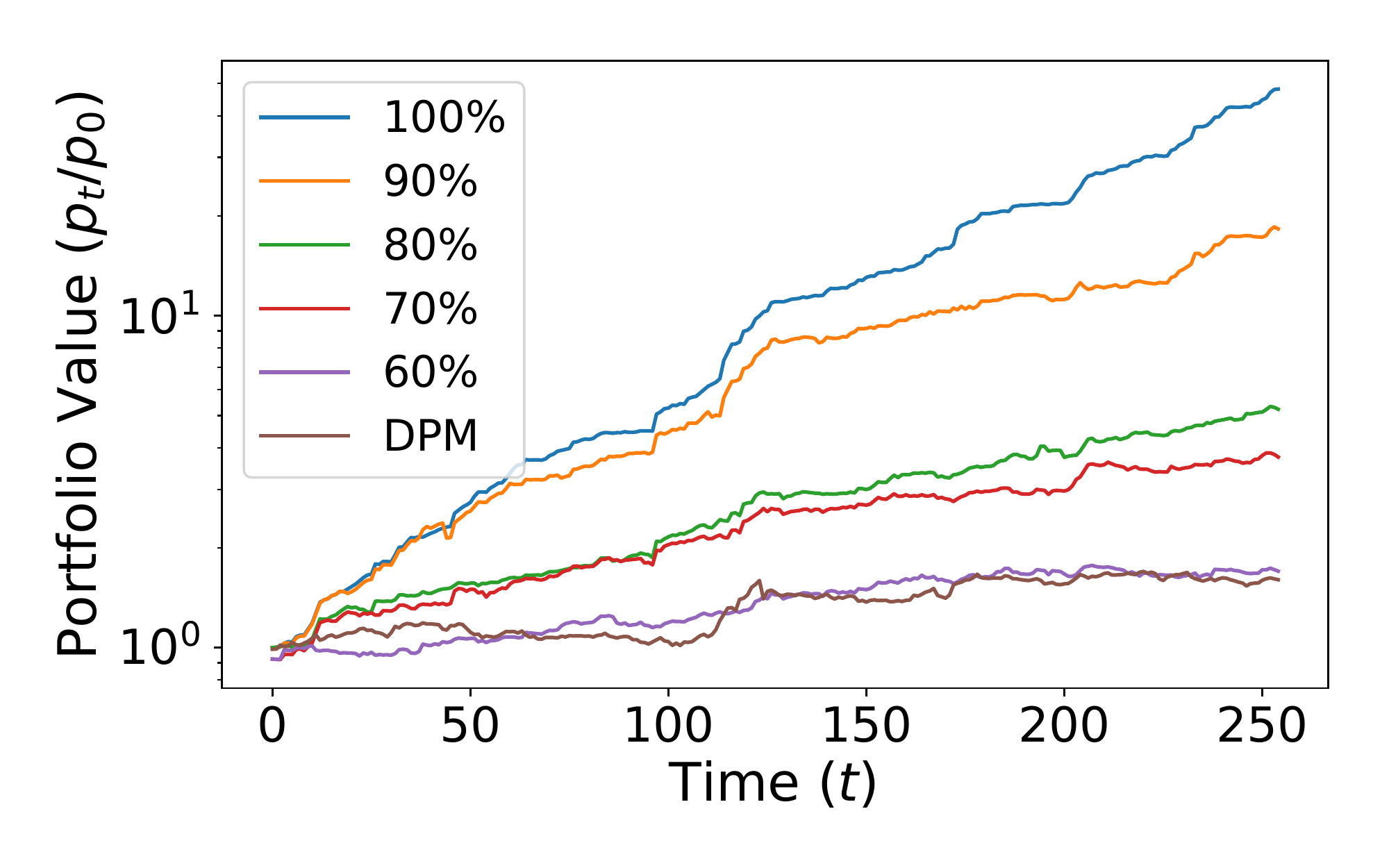}
        \caption{HighTech simulation}
        \label{fig:reuters_simulation}
    \end{subfigure}
    \caption{The simulation results of labels with different accuracy on Bitcoin and HighTech dataset. 
    SARL's performance increases with the accuracy and it outperforms DPM even with low (60\%) accuracy. 
    }
    \label{fig:simulation}
    \end{figure}

    \begin{figure}[t!]
    \centering
    \includegraphics[width=1\columnwidth]{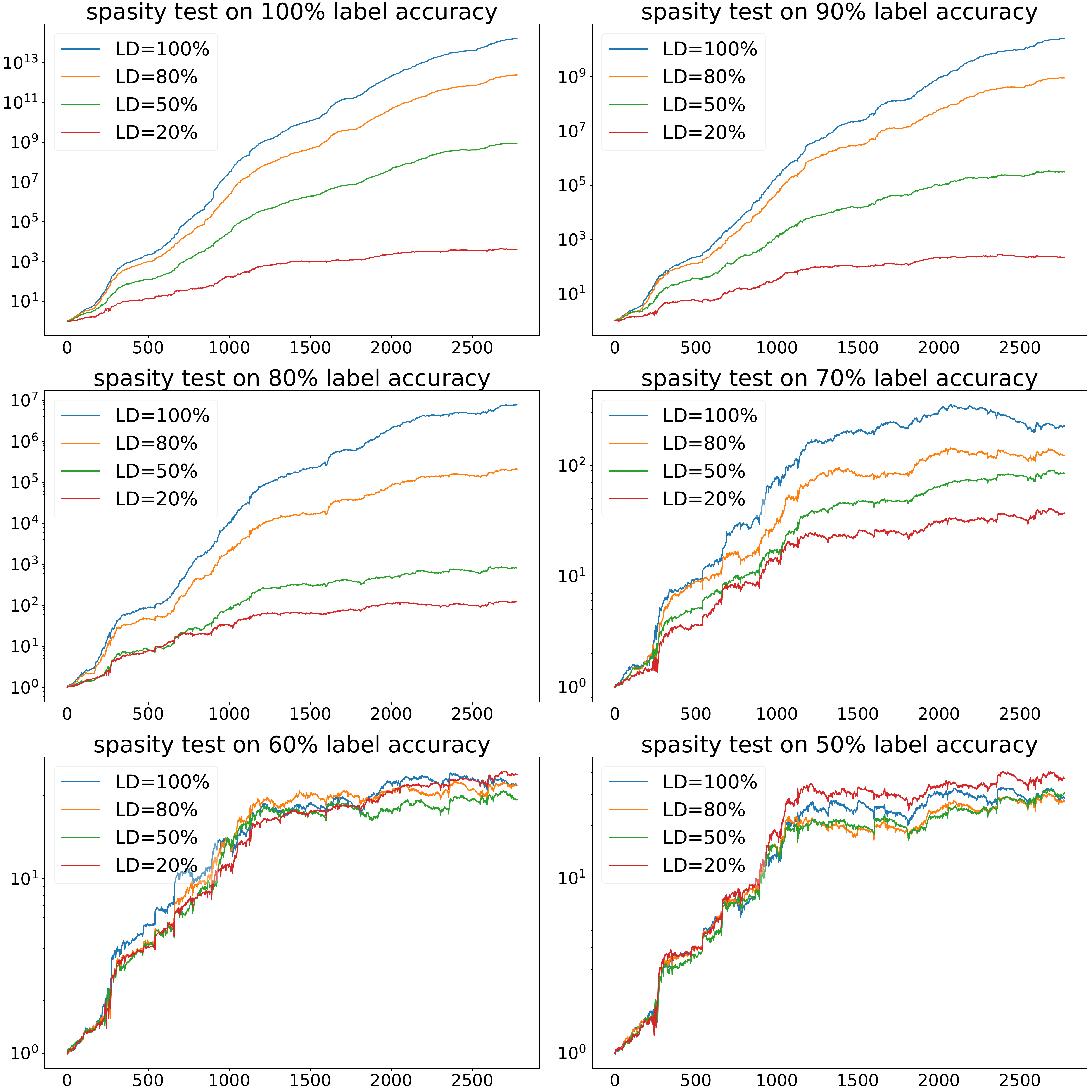}
    \caption{The sparsity test on Bitcoin dataset. Each subfigure shows the results of SARL whose state is augmented by simulated price prediction labels with different sparsity. LD is label density from $\{20\%, 50\%, 80\%, 100\%\}$.}
    \label{fig:bitcoin_sparsity}
    \end{figure}
    
\textbf{Sharpe Ratio (SR).} We report the SR of different time period in Table~\ref{tab:Bitcoin} and \ref{tab:Reuters}. In PM, SR often decreases over time in testing set due to the environmental distribution shift. Although the training objective function of SARL does not involve a risk-related concept, SARL attains the highest SR in most cases, suggesting its ability to learn risk-adverse PM strategies.
Moreover, the advantage of SARL over DPM also validates the importance of incorporating external features for PM, as it enhances the robustness of the agent.

    \begin{figure}[t]
    \centering
    \includegraphics[width=0.95\columnwidth]{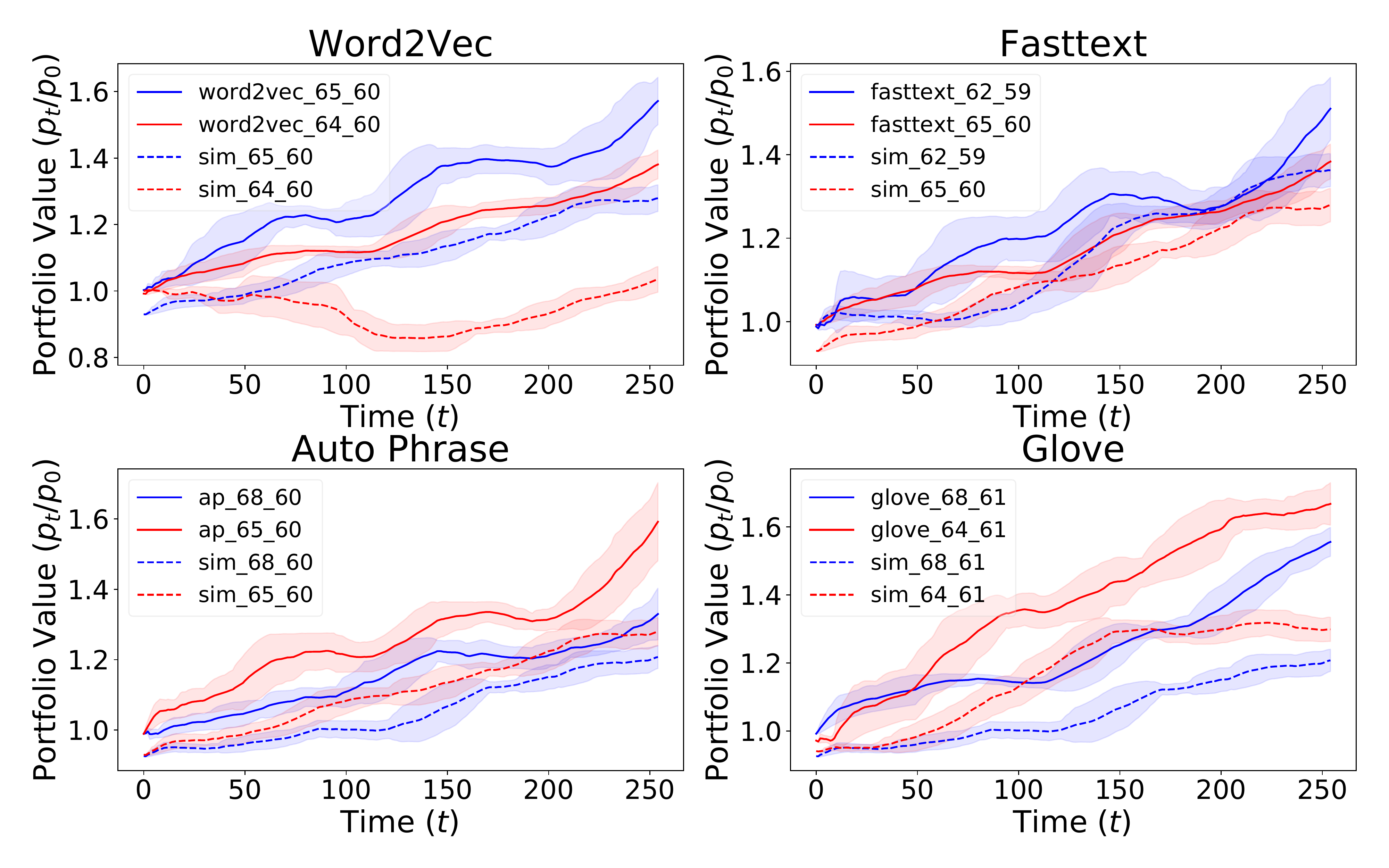}
    \caption{The comparison between states augmented by different news embedding and randomly simulated labels.}
    \label{fig:correlation}
    \end{figure}

\textbf{Studies on prediction accuracy and sparsity.} 
Here we use simulated prediction labels as state augmentation to study the effect of prediction accuracy and label density (frequency of having label prediction per each time step) on PM.
Using the same label density pattern as the original datasets, Figure~\ref{fig:simulation} 
shows that when the accuracy is higher than $60\%$, albeit noisy prediction, SARL can learn better PM strategy. The higher the accuracy is, the better the strategy is. In addition, we also run simulations for the sparsity test when varying label density. 
Figure~\ref{fig:bitcoin_sparsity} shows that the performance of PM on the Bitcoin dataset also benefits from higher label density.  
With the simulation result, we can find that even if we only have labels with density $50\%$ but $70\%$ accuracy, the portfolio value still has an improvement of $90.3\%$ when compared to DPM. This phenomenon suggests that if we have labels with high confidence, we can improve the PM strategy even if the labels are sparse. Similar observations also hold for the HighTech dataset, where the figure of the sparsity test is given in Appendix.

\textbf{Hidden Correlation in News.} The simulation result in Figure~\ref{fig:simulation} shows relatively mild improvement of SARL over DPM even when the accuracy is $60\%$.
However, when we use the embedding of the financial news to augment the state, we obverse significant improvement with similar (around $60\%$) accuracy in testing set. The results reveal the hidden correlation in financial news as opposed to randomly simulated label predictions. 
To validate the hidden correlation from financial news that can be exploited by SARL, we use different word embedding methods as in Table \ref{tab:NLP} and train SARL with the same simulated training/testing accuracy on the HighTech dataset for comparison.
Figure~\ref{fig:correlation} verifies the existence of the hidden correlation in financial news. The results
show that the performance of SARL when trained on financial news is much better than that when trained on randomly simulated labels with the same accuracy. The results also corroborate that SARL is capable of learning from external information to improve PM.

\textbf{Training SARL with different RL algorithms.} As our SARL framework is compatible with standard RL training, we employ different RL algorithms including DPG~\cite{silver2014deterministic}, PPO~\cite{schulman2017proximal}, PG~\cite{sutton2000policy} to verify the generality of SARL. Most of SARL variants across different RL training methods indeed show consistent performance gains over DPM.  We refer the readers to the Appendix for detailed comparisons.

\section{Conclusion}
In this paper, we propose SARL, a novel and generic state-augmented RL framework that can integrate heterogeneous data sources into standard RL training pipelines for learning PM strategies. Tested on the Bitcoin and Hightech datasets, SARL can achieve significantly better portfolio value and Sharpe ratio. We also conducted comparative experiments and extensive simulations to validate the superior performance of SARL in PM. 
We believe this work will shed light on more effective and generic RL-based PM algorithms.

\section{Acknowledgements}
This work was supported by National Key Research and Development Program of China (2018AAA0101900), Zhejiang Natural Science Foundation (LR19F020002, LZ17F020001), National Natural Science Foundation of China(61976185, U19B200023, 61572431), the Fundamental Research Funds for the Central Universities and Chinese Knowledge Center for Engineering Sciences and Technology, and IBM-ILLINOIS Center for Cognitive Computing Systems Research (C3SR) -- a research collaboration as part of the IBM AI Horizons Network, National Science Foundation award CCF-1910100 and DARPA award ASED-00009970.
\bigskip
\bibliographystyle{aaai}
\bibliography{AAAI.4483.bib}
\newpage
\section{Appendix}
\subsection{Compatibility of SARL}
\begin{figure}[h]
\centering
\includegraphics[width=0.5\textwidth]{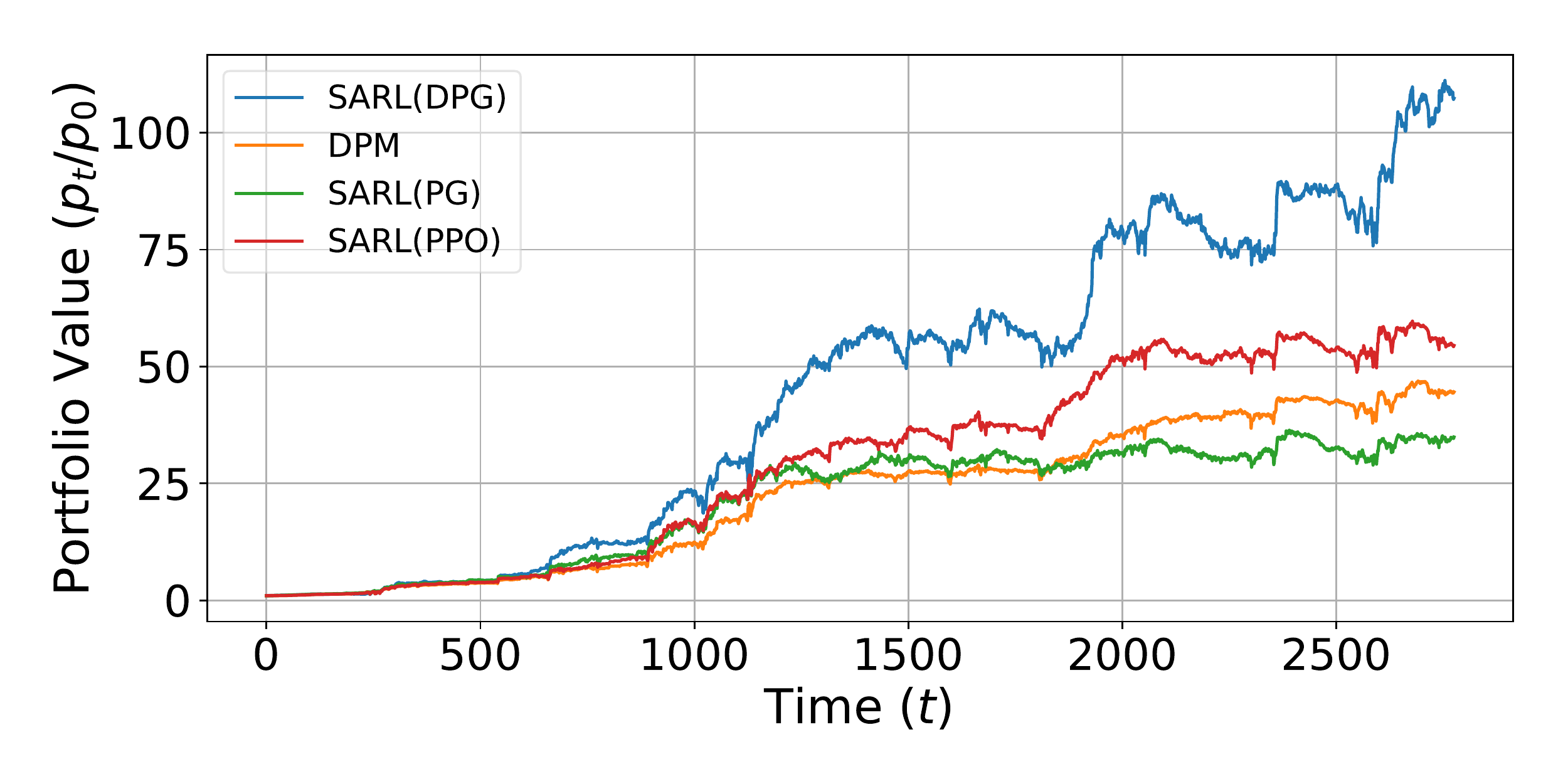}
\caption{The comparison among different RL algorithms with our SARL framework on Bitcoin dataset}
\label{fig:bitcoin_comp}
\end{figure}

\begin{figure}[!h]
\centering
\includegraphics[width=0.5\textwidth]{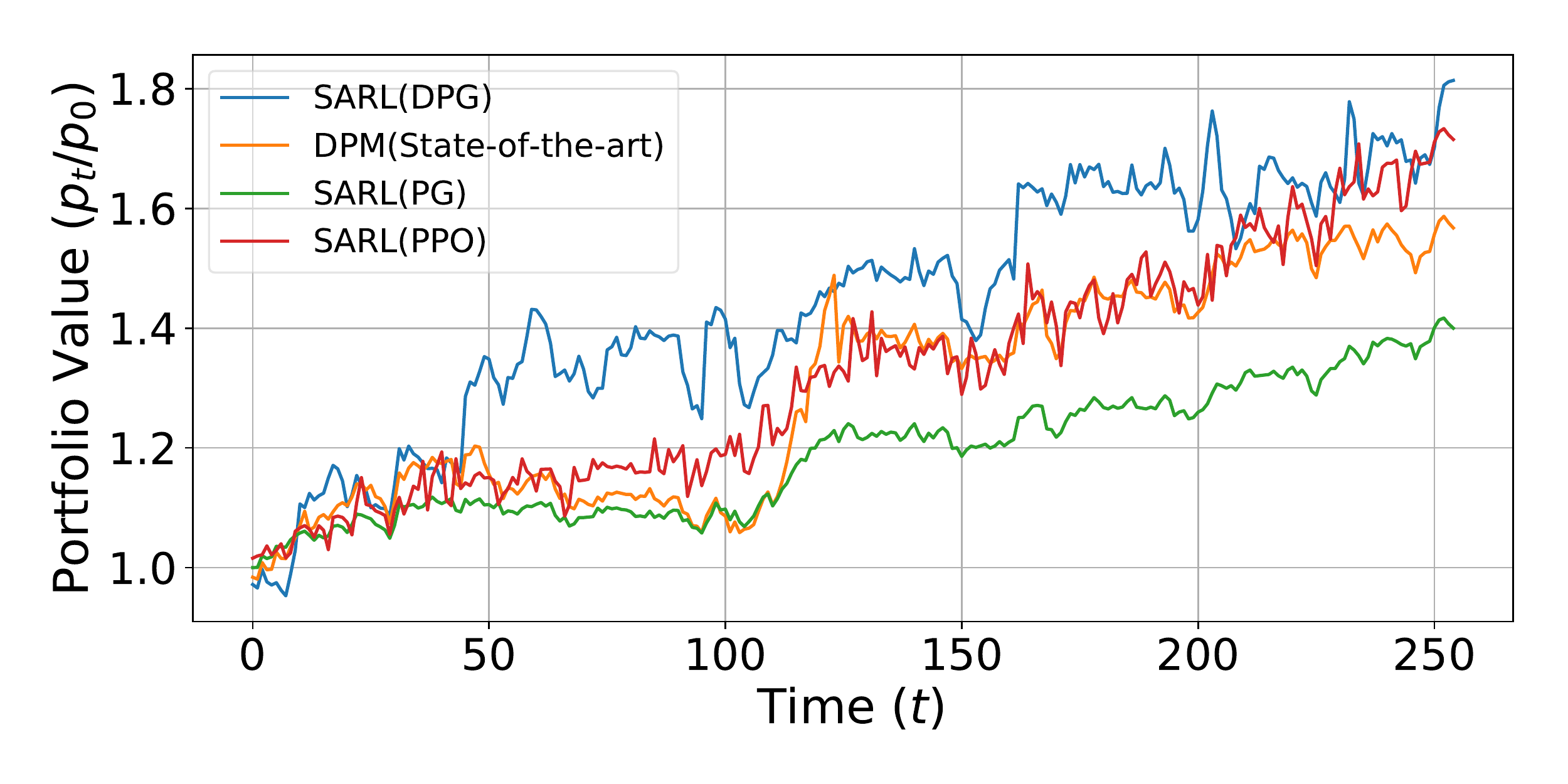}
\caption{The comparison among different RL algorithms with our SARL framework on HighTech dataset}
\label{fig:reuters_comp}
\end{figure}

\begin{table}[!h]
\centering
\caption{Sharpe Ratio of different time period on Bitcoin dataset. (w:week, m:month)}
\begin{tabular}{c|cccc}
\hline
\textbf{sharpe ratio} & \textbf{1w}    & \textbf{2w}    & \textbf{1m}    & \textbf{2m}    \\ \hline
\textbf{DPM}          & 14.25          & 13.37          & 12.64          & 9.80           \\ \hline
\textbf{SARL(PG)}     & 14.40          & 12.51          & 12.24          & 9.51           \\ \hline
\textbf{SARL(PPO)}    & 13.94          & 12.66          & 12.67          & 10.21          \\ \hline
\textbf{SARL(DPG)}    & \textbf{16.66} & \textbf{15.52} & \textbf{13.50} & \textbf{10.25} \\ \hline
\end{tabular}
\end{table}

\begin{table}[!h]
\centering
\caption{Sharpe Ratio of different time period on HighTech dataset. (w:week, m:month)}
\begin{tabular}{c|ccccc}
\hline
\textbf{sharpe ratio} & \textbf{1w}   & \textbf{2w}   & \textbf{1m}   & \textbf{3m}   & \textbf{6m}   \\ \hline
\textbf{DPM}          & 2.15          & 6.44          & 3.88          & 2.31          & 2.22          \\ \hline
\textbf{SARL(PG)}     & 8.93          & 6.47          & \textbf{6.44} & 2.03          & 1.65          \\ \hline
\textbf{SARL(PPO)}    & \textbf{9.44} & 4.13          & 2.26          & 1.39          & 1.61          \\ \hline
\textbf{SARL(DPG)}    & 7.69          & \textbf{7.73} & 3.83          & \textbf{2.91} & \textbf{2.37} \\ \hline
\end{tabular}
\end{table}

\newpage
\subsection{Sparsity of HighTech}
\begin{figure}[h]
\centering
\includegraphics[width=0.5\textwidth]{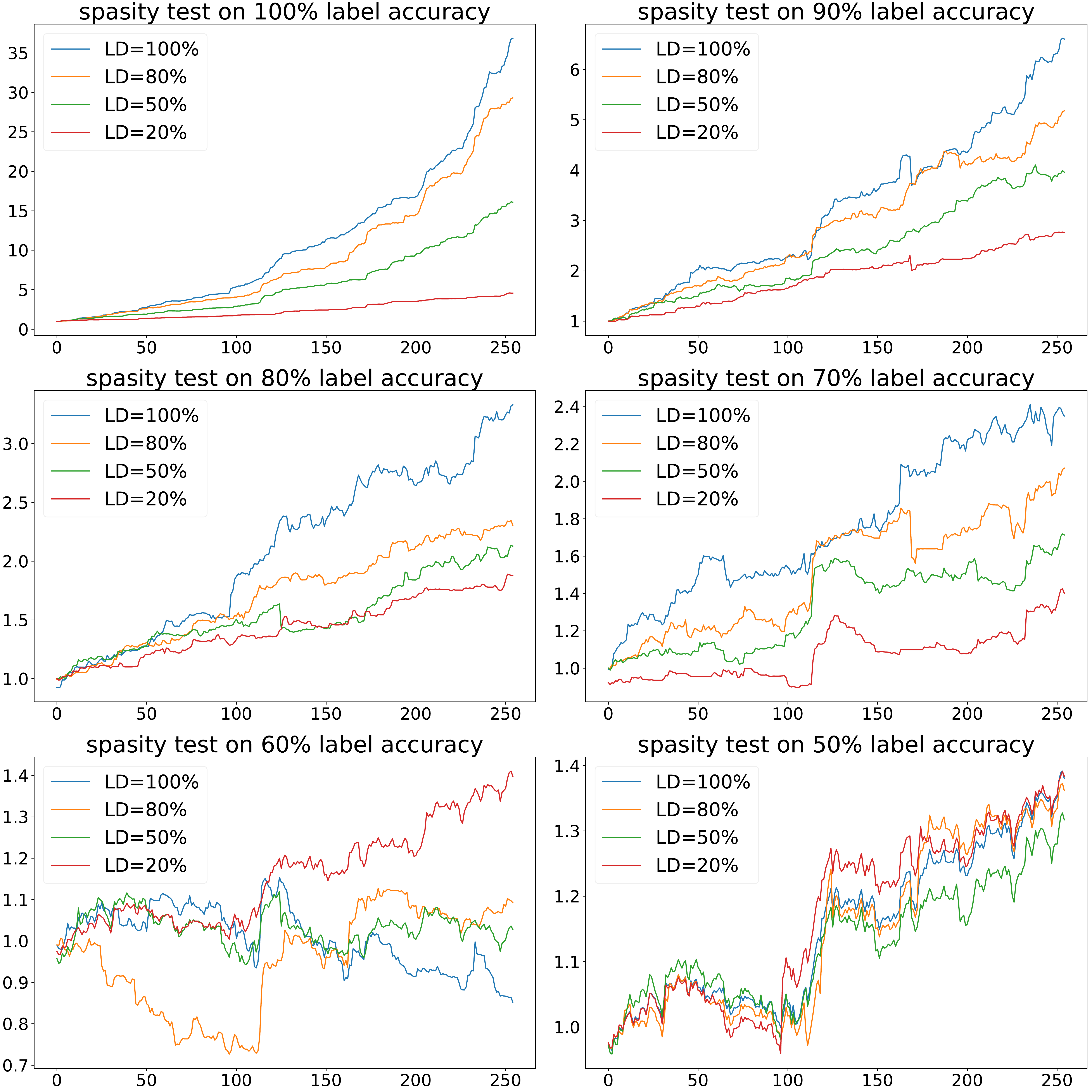}
\caption{The sparsity test on HighTech dataset. Each subfigure shows the results of SARL whose state is augmented by financial news embedding with different sparsity. LD is label density from $\{20\%, 50\%, 80\%, 100\%\}$.}
\label{fig:reuters_sparsity}
\end{figure}
\end{document}